\documentclass[aps,prl,twocolumn,amsmath,amssymb,superscriptaddress,reprint,floatfix]{revtex4-1}

\usepackage[utf8]{inputenc}
\usepackage[colorlinks=true,citecolor=blue,urlcolor=blue]{hyperref}
\usepackage{graphicx}
\usepackage{xcolor}
\usepackage{units}
\usepackage{mhchem}
\usepackage{dblfloatfix}
\usepackage{soul}

\makeatother

\begin{document}

\title{Experimental realization of a quantum dot energy harvester}

\author{G. Jaliel}
\altaffiliation{jg737@cam.ac.uk}
\affiliation{Cavendish Laboratory, University of Cambridge, JJ Thomson Avenue, Cambridge CB3 0HE, United Kingdom}
\author{R.~K. Puddy}
\affiliation{Cavendish Laboratory, University of Cambridge, JJ Thomson Avenue, Cambridge CB3 0HE, United Kingdom}
\author{R. S\'{a}nchez}
\affiliation{Departamento de F\'isica Te\'orica de la Materia Condensada and Condensed Matter Physicas Center (IFIMAC), Universidad Aut\'onoma de Madrid, E-28049 Madrid, Spain}
\author{A.~N. Jordan}
\affiliation{Department of Physics and Astronomy, University of Rochester, Rochester, New York 14627, USA}
\author{B. Sothmann}
\affiliation{Theoretische Physik, Universit\"at Duisburg-Essen and CENIDE, D-47048 Duisburg, Germany}
\author{I. Farrer}
\affiliation{Department of Electronic and Electrical Engineering, University of Sheffield, Mappin Street, Sheffield  S1 3JD, United Kingdom}
\author{J.~P. Griffiths}
\affiliation{Cavendish Laboratory, University of Cambridge, JJ Thomson Avenue, Cambridge CB3 0HE, United Kingdom}
\author{D.~A.  Ritchie}
\affiliation{Cavendish Laboratory, University of Cambridge, JJ Thomson Avenue, Cambridge CB3 0HE, United Kingdom}
\author{C.~G. Smith}
\affiliation{Cavendish Laboratory, University of Cambridge, JJ Thomson Avenue, Cambridge CB3 0HE, United Kingdom}

\begin{abstract}
We demonstrate experimentally an autonomous nanoscale energy harvester that utilises the physics of resonant tunnelling quantum dots. Gate defined quantum dots on \ce{GaAs/AlGaAs} high-electron-mobility transistors are placed on either side of a hot electron reservoir. The discrete energy levels of the quantum dots are tuned to be aligned with low energy electrons on one side and high energy electrons on the other side of the hot reservoir. The quantum dots thus act as energy filters and allow for the conversion of heat from the cavity into  electrical power. Our energy harvester, measured at an estimated base temperature of \unit[75]{mK} in a He$^3$/He$^4$ dilution refrigerator, can generate a thermal power of \unit[0.13]{fW} for a temperature difference across each dot of about \unit[67]{mK}. 
\end{abstract}

\maketitle
In recent years there has been an increased interest in devices which can convert waste heat into useful work~\cite{white_energy-harvesting_2008}. Thermoelectric generators 
where a temperature bias applied to an electric conductor gives rise to a charge current flow  are good candidates~\cite{shakouri_recent_2011,benenti_fundamental_2017}.
Unfortunately, current thermoelectric devices have relatively small efficiencies~\cite{rowe_thermoelectric_2006}.
This issue can be overcome by nanoscale thermoelectrics where engineered bandstructures and quantum mechanical effects can give rise to an increased efficiency~\cite{hicks_effect_1993,hicks_thermoelectric_1993,mahan_best_1996}.
Quantum dots constitute an important element in designing highly efficient thermoelectrics~\cite{beenakker_theory_1992,humphrey_reversible_2002,nakpathomkun_thermoelectric_2010,esposito_thermoelectric_2009} because their discrete resonant levels provide excellent energy filters. Thermoelectric effects have been investigated in various quantum-dot setups~\cite{staring_coulomb-blockade_1993,dzurak_observation_1993,dzurak_thermoelectric_1997,godijn_thermopower_1999,small_modulation_2003,llaguno_observation_2003,scheibner_thermopower_2005,scheibner_sequential_2007,svensson_lineshape_2012,svensson_nonlinear_2013,thierschmann_diffusion_2013,josefsson_quantum-dot_2018}.

Energy harvesting devices require that the energy source is separated from the electrical circuit, so no charge is extracted from it~\cite{sothmann_thermoelectric_2015}. This can be accomplished in three-terminal devices where a hot terminal injects heat but no charge into the setup, thus driving a charge current between two cold reservoirs. There have been a number of proposals for these kinds of energy harvesters~\cite{entin-wohlman_three-terminal_2010,sanchez_optimal_2011,sothmann_rectification_2012,sothmann_magnon-driven_2012,ruokola_theory_2012,jiang_thermoelectric_2012,jiang_three-terminal_2013,machon_nonlocal_2013,jordan_powerful_2013,sothmann_powerful_2013,bergenfeldt_hybrid_2014,donsa_double_2014,mazza_thermoelectric_2014,mazza_separation_2015,sanchez_chiral_2015,sanchez_heat_2015,hofer_quantum_2015,bosisio_nanowire-based_2016,szukiewicz_optimisation_2016,jiang_near-field_2018}.
Three-terminal heat engines based on Coulomb-coupled quantum dots~\cite{sanchez_optimal_2011,sothmann_rectification_2012} have been realized experimentally recently~\cite{thierschmann_three-terminal_2015,roche_harvesting_2015,hartmann_voltage_2015}. Due to their design they are however limited to low power. A three-terminal energy harvester based on two resonant-tunneling quantum dots with different energy levels overcomes this issue. It can in principle reach Carnot efficiency and can be optimized to achieve a large power in combination with a high efficiency at maximum power~\cite{jordan_powerful_2013}.
\begin{figure}[b]
   \includegraphics[width=0.82\linewidth]{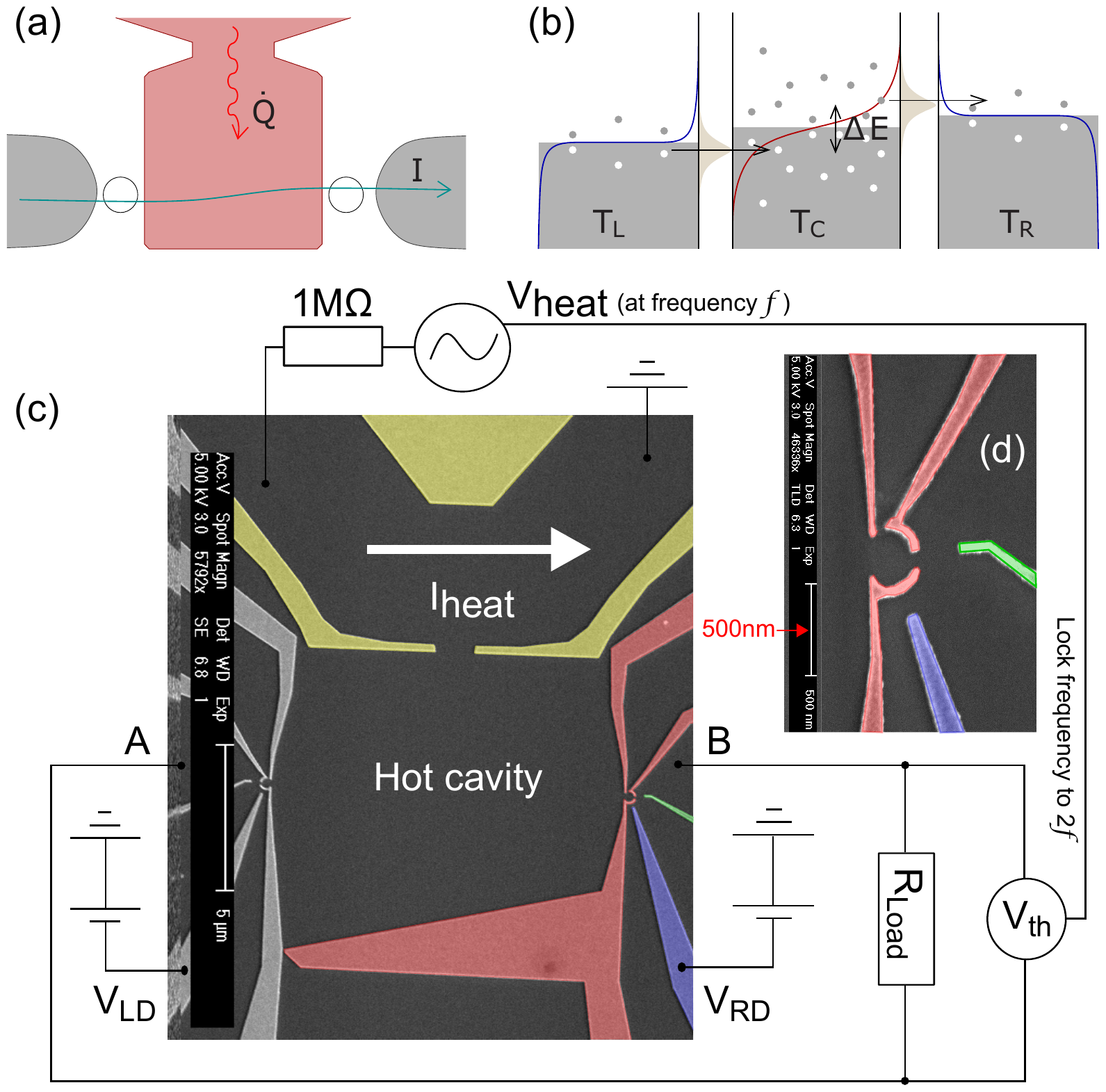}
   \caption{Resonant tunneling energy harvester. (a) Two quantum dots connect two electronic leads (at temperature $T_0=T_{\rm L}=T_{\rm R}$) to a hot cavity at $T_{\rm C}$. A heat current $\dot Q$ at frequency $f$ is absorbed by the flowing electrons to generate a heat current $I$ at frequency $2f$. (b) Relative energy diagram of the heat engine. Tuning the resonant level positions filters tunneling transitions with an energy gain $\Delta E$. (c) False colour SEM image of the device with the electrical circuit used for the thermopower measurement. (d) shows a zoomed false-colour-SEM image of the right quantum dot. }
   \hfill\null
   \label{model}
\end{figure}
A similar device has also been proposed~\cite{edwards_quantum-dot_1993,edwards_cryogenic_1995}, and later demonstrated~\cite{prance_electronic_2009}, as a building block of a nanoscale refrigerator. In this manuscript, we experimentally realize a resonant-tunneling energy harvester and demonstrate its ability to generate electrical power in an external load arising from energy exchanges between a hot and a cold reservoir. Importantly, no external drive or cycling is required; that is, the system is entirely autonomous and begins producing power as soon as a thermal gradient is present.

The system we have investigated, shown in Fig.~\ref{model}, is comprised of two quantum dots that connect a hot cavity to two cold reservoirs \cite{jordan_powerful_2013}. By putting two quantum dots in series with a hot cavity, electrons that enter via the left dot are forced to gain a prescribed energy in order to exit through the right dot, transporting a single electron charge from left to right, cf. Fig.~\ref{model}(a). Constrained by the conservation of global charge and energy in the device, this thermal energy gain of electrons will be converted into electrical current \cite{jordan_powerful_2013}.
 
Fig.~\ref{model}(c) shows a false-colored scanning electron micrograph (SEM) image of a typical device we tested, along with the electrical circuit used in the experiments. \ce{Ti/Au} gates were patterned on the surface of \ce{GaAs/AlGaAs} heterostructure material using electron-beam lithography. The 2DEG was \unit[110]{nm} below the surface, and was contacted by annealing \ce{AuGeNi} ohmic contacts. The mobility $\mu$ and carrier concentration $n$ of the 2DEG were measured to be $\mu \approx \unit[3.38\times 10^6]{cm^2V^{-1}s^{-1}}$ and $n \approx \unit[1.35\times 10^{11}]{cm^{-2}}$ at \unit[1.5]{K}. The surface gates define a cavity of $\unit[90]{\mu  m^2}$ area at the central 2DEG region with two quantum dots respectively on the left and the right sides, and a heating channel on the top. The quantum dots, of \unit[310]{nm} diameter, as shown in \ref{model}(d), are constructed of three barrier gates (colored red in Fig.~\ref{model}(c)), one detector gate (colored green), and one plunger gate (colored blue). Both dots were found to have charging energies of approximately \unit[1.5]{meV} and first excited states always at least $\unit[250]{\mu eV}$ above the ground state. The top heating channel (gates colored yellow in Fig.~\ref{model}(c)) is connected to the central cavity via a gap of $\unit[1.26]{\mu m}$, which allows hot electrons to traverse into the cavity and form different temperature profiles.  Measurements were performed in a He\textsuperscript{3}/He\textsuperscript{4} dilution refrigerator at an estimated base temperature $T_0$ of \unit[75]{mK}. The experiment was repeated with two samples, which are similar in design but have different resonances for the quantum dots. 

The thermal power generated by the energy harvester was measured with the set-up in Fig.~\ref{model}(c). An AC current $I_\text{Heat}$ which heats electrons at frequency $f=\unit[33]{Hz}$ was applied to the heating channel using a lock-in amplifier, and the thermal voltage $V_\text{th}$ was measured across \emph{A - B}, with another amplifier locking in at frequency 2\emph{f}, whilst stepping the voltage $V_\text{LD}$ on the left dot plunger gate and sweeping the voltage $V_\text{RD}$ on the right dot plunger gate through their respective Coulomb resonance. Since the heating power varies as $I^2$, the electron temperature in the cavity oscillates at twice the frequency of the current $I_\text{Heat}$. Thus it was necessary to phase lock to the 2\emph{f} component of $V\textsubscript{th}$ \cite{dzurak_thermoelectric_1997}. 
The temperatures of the central cavity for different AC currents may be estimated by fitting the differential conductance of the quantum dots with a thermally broadened resonance, see Supplementary Information for details. The cold reservoirs, which we assume to be at base temperature $T_0$, are connected externally by a load resistor, $R\textsubscript{Load}$. The thermal power is then extracted by $P=V_{\rm th}^2/R\textsubscript{Load}$. In potential applications the heating channel would be replaced by the heat source we wish to harvest energy from and the $R\textsubscript{Load}$ represents an external device where useful work is done~\cite{jordan_powerful_2013}. 

\begin{figure}[t]
    \centering
    \includegraphics[width=1.0\linewidth]{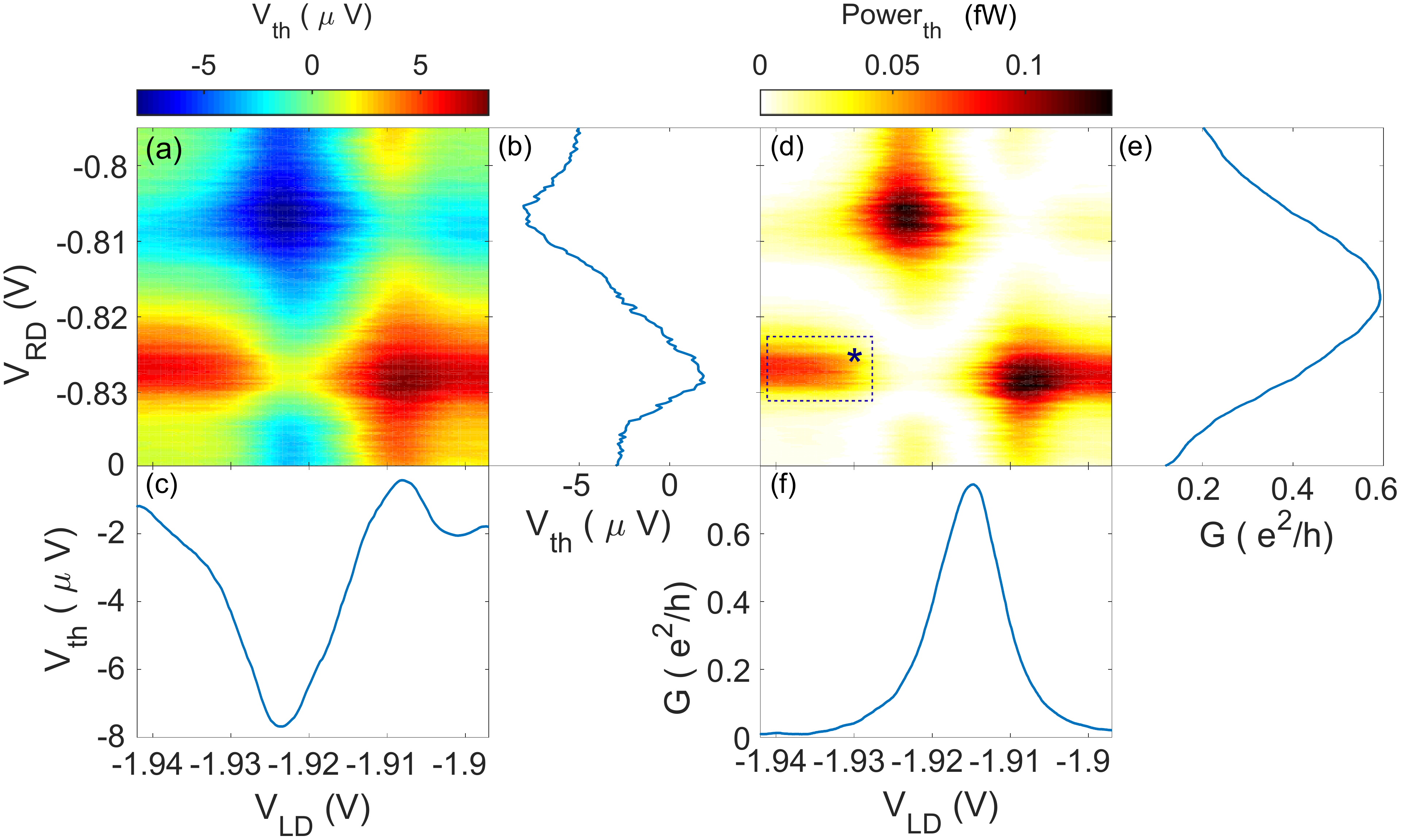}
    \caption{(a) The thermal voltage V\textsubscript{th}, across the device, as a function of left and right plunger gates measured whilst an AC current, I\textsubscript{heat} = $\unit[100]{nA}$, is applied to the heating channel. The applied I\textsubscript{heat} results in an estimated temperature difference of $\Delta T= T_{C}-T_{L} \approx \unit[47]{mK}$ across the dots. (b) and (c) line graphs through (a) at $V\textsubscript{LD} = \unit[-1.924]{V}$ and $V\textsubscript{RD} = \unit[-0.805]{V}$ respectively. (d) Estimated power output of the device showing the two expected operational points and a third (highlighted by the box with a mark star $*$) due to external circuit impedance. The power is given by $P=V_{\rm th}^2/R_\text{Load}$ where $R_\text{Load}$ is the resistance loading on the circuit. (e, f) Conductance peaks of the two dots as a function of the respective gate voltage.}
    \label{Vth01}
\end{figure}

Figure~\ref{Vth01}(a) shows the thermal voltage between A and B (in Fig.\ref{model}(c)), $V_\text{th}$, measured with a heating current $I_\text{Heat}=\unit[100]{nA}$ and a load resistor $R_\text{Load}=\unit[500]{k\Omega}$ in the circuit. The negative thermal voltage appears at $V_\text{LD}\approx\unit[-1.924]{V}$ for the left dot, as shown in Fig.~\ref{Vth01}(c), and $V_\text{RD}\approx\unit[-0.805]{V}$ for the right dot, as in Fig.~\ref{Vth01}(b). Fig.~\ref{Vth01}(d) is the thermal power extracted from the thermal voltage of Fig.\ref{Vth01}(a), through $P=V_\text{th}^2/R\textsubscript{Load}$. The maximum thermal power is found at  $\unit[(-1.924,-0.805)]{V}$, followed by the second largest thermal power point at $\unit[(-1.907,-0.829)]{V}$, in Fig.~\ref{Vth01}(d). The maxima appear in the vicinity of the electrical conductance peaks shown in Fig.~\ref{Vth01}(e) and Fig.~\ref{Vth01}(f) respectively. This is because when both charge and heat are exclusively carried by electrons, for both diffusive and ballistic transport, the Seebeck coefficient (thermopower) $S$ is related to the energy derivative of the conductance  $G$ \cite{appleyard_thermometer_1998},
\begin{equation}
     S = \left. \frac{V_{th}}{T_{C}-T_0} \right\vert_\text{I=0} = -\frac{\pi^2k_{B}^2}{3e}(T_{C}-T_0) \frac{\partial \ln G}{\partial \mu}. 
     \label{thermopower}
\end{equation}
Here $T_{C}$ is the electron temperature of the cavity, $T_0$ is the temperature of cold reservoirs, and $\mu$ is the chemical potential of the contacts. Meanwhile, the thermal voltage peaks in Fig.~\ref{Vth01}(a) are also related to the energy derivative of the conductance of the Fig.~\ref{Vth01}(e) and (f),  as shown Eq.~\eqref{thermopower}. Some thermopower is detected while only one dot is open on resonance, such as the area marked by the star $\star$ in Fig.~\ref{Vth01}(d). This arises from the influence of the external circuit impedance. 

\begin{figure}[t]
    \centering
    \includegraphics[width=0.95\linewidth]{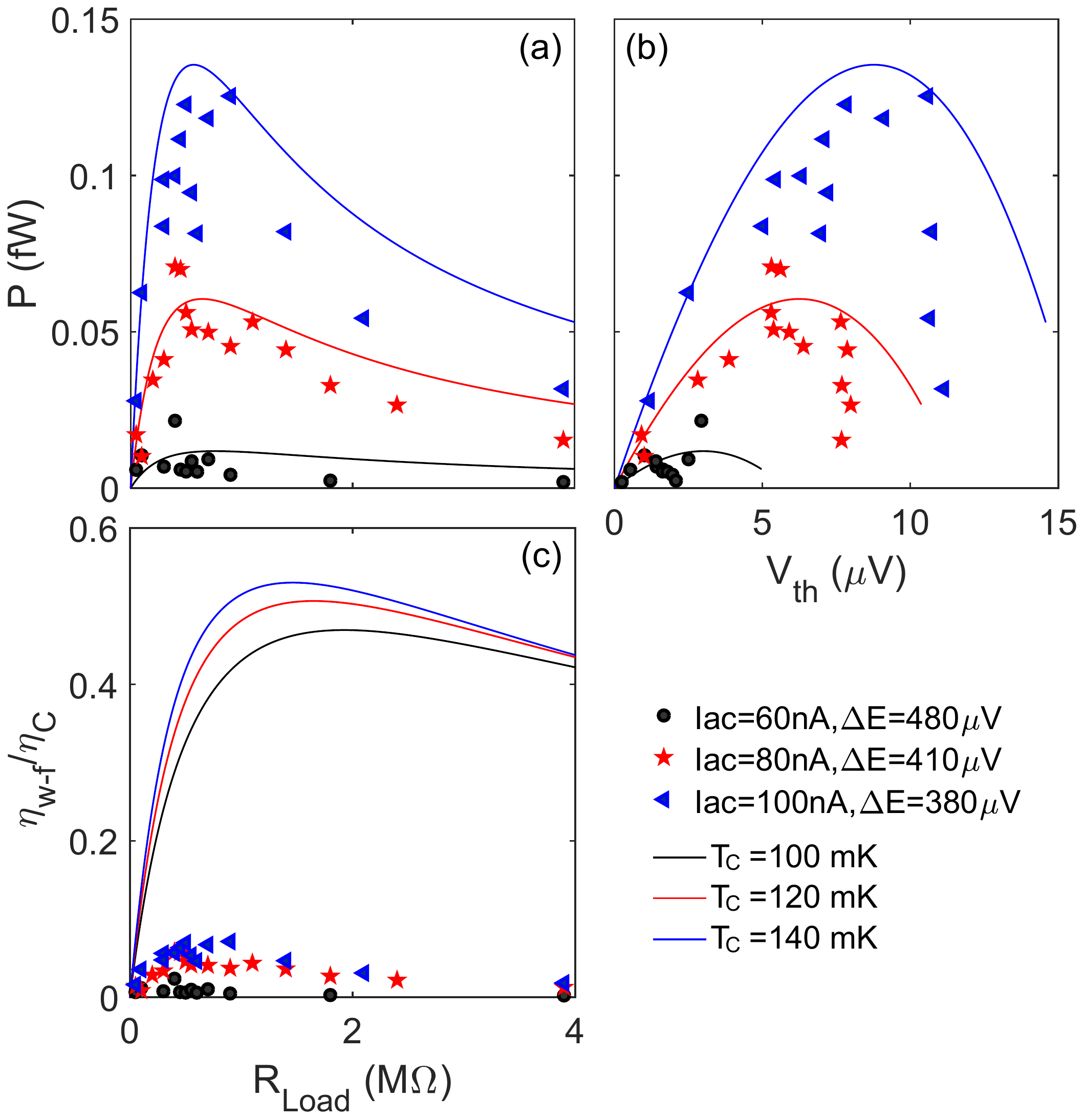}
    \caption{ Engine characteristics. The points of black circles, red stars and blue triangles show results of experimental measurements. Panel (a) depicts the maximum thermal power from the measurements with different loading resistance, whilst applying AC current $\unit[60]{nA}$ (black circles), $\unit[80]{nA}$ (red stars) and $\unit[100]{nA}$ (blue triangles) on the heating channel. Panel (b) shows the thermal power and its relative thermal voltage, which is also the bias voltage between $A-B$ in Fig.\ref{model}(c). Panel (c) depicts the ratio of the estimated efficiency through Eq.~\eqref{efficiency} with the Carnot efficiency while changing the resistance. The solid lines show the relative theoretical modeling for different heating currents leading to different cavity temperatures, $T_{\rm C}$. The theoretical efficiency in (c) is computed from Eq.~\eqref{eq:theff}, with resonances of $\Gamma_{\rm L}=\Gamma_{\rm R}=\unit[3.5]{\mu eV}$ and the energy level difference of $\Delta E=\unit[45]{\mu eV}$ of the two dots. Parameter $A=0.8$ is related to the quantum dot barriers. The base temperature in the theoretical model is $\unit[85]{mK}$. }
    \label{VandP}
\end{figure}

Thermopower measurements were carried out using resistance values ($R\textsubscript{Load}$) from $\unit[50]{k\Omega}$ to $\unit[3.9]{M\Omega}$ in the circuit, whilst an AC current of $\unit[60]{nA}$, $\unit[80]{nA}$ and $\unit[100]{nA}$ is applied on the heating channel. The heating currents of $\unit[60]{nA}$, $\unit[80]{nA}$ and $\unit[100]{nA}$ correspond to $\unit[122]{mK}$, $\unit[130]{mK}$ and $\unit[140]{mK}$ respectively, as discussed in Supplementary Material. Figure~\ref{VandP} depicts the maximal generated power for each measurement as a function of the load resistance and the relative thermal voltages respectively, where black circles represent the experimental data while a current of $\unit[60]{nA}$ is applied on the heating channel, red stars for $\unit[80]{nA}$, and blue triangles for $\unit[100]{nA}$. (Solid lines represent results from theoretical modeling and will be discussed later.) For increasing resistance $R_{\rm Load}$, the power increases, reaches a maximum and then drops down, as shown in Fig.~\ref{VandP}(a). As the heating current in the channel is increased, the power also rises. This is because the cavity temperature increases with the heating current, resulting in more electrons tunneling through the two dots and converting more energy to electrical current efficiently, as predicted in the theoretical proposal \citep{jordan_powerful_2013}. Interestingly, the maximum power always appears around $R_{\rm Load}\approx \unit[500]{k\Omega}$ for all heating currents, corresponding to impedance matching between the heat engine and the resistor. The power vs. thermal voltage in Fig.~\ref{VandP}(b) gives an estimation of the open-circuit stall voltage of our device in each configuration. In the linear regime one expects the maximal power to occur at half the stall voltage~\cite{sothmann_thermoelectric_2015}. The asymmetric dependence of the measurements suggests the presence of non-linear effects.

We next turn to the efficiency of heat to work conversion which is defined as the ratio of the generated electrical power $P$ to the heat current from the hot reservoir ${\dot{Q}}$. The heat current is given by $\dot{Q} = \kappa \Delta T =\kappa(T_{\rm C}-T_0)$ where the thermal conductance $\kappa$ can be estimated from the electrical conductance $G$ via the Wiedemann-Franz law, $\kappa=G L \overline{T}$, where $L$ is the Lorenz number, $\overline{T} = (T_{\rm C}+T_0)/2$ and $G$ is the combination of the conductance at $V_{\rm RD}=\unit[-0.805]{V}$ in Fig.~\ref{Vth01}(e) and that at $V_{\rm LD}=\unit[-1.924]{V}$ in Fig.~\ref{Vth01}(f). 
We remark that the Wiedemann-Franz law in general is violated for mesoscopic conductors with strongly energy dependent transmissions such as quantum dots~\cite{kubala_violation_2008, tsaousidou_thermoelectric_2010,erdman_thermoelectric_2017,dutta_thermal_2017}. As the thermal conductance cannot be measured directly in our setup, we still use it to obtain an lower bound on the thermoelectric efficiency given by
\begin{equation}
\eta_\text{w-f}=\frac{V_{th}^2}{\kappa \Delta T R_{\rm Load}}=\frac{V_{th}^2}{G L \overline{T} \Delta TR_{\rm Load}},
\label{efficiency}
\end{equation}
which can be compared with the theoretical efficiency calculated below. Figure~\ref{VandP}(c) depicts the ratio of the estimated efficiency from Equation (\ref{efficiency}) to the Carnot efficiency ($\eta_C=1-T_{\rm 0}/T_{\rm C}$) for $\unit[60]{nA}$, $\unit[80]{nA}$ and $\unit[100]{nA}$ on the heating channel respectively.

\begin{figure}[ht]
    \centering
 \includegraphics[width=0.95\linewidth]{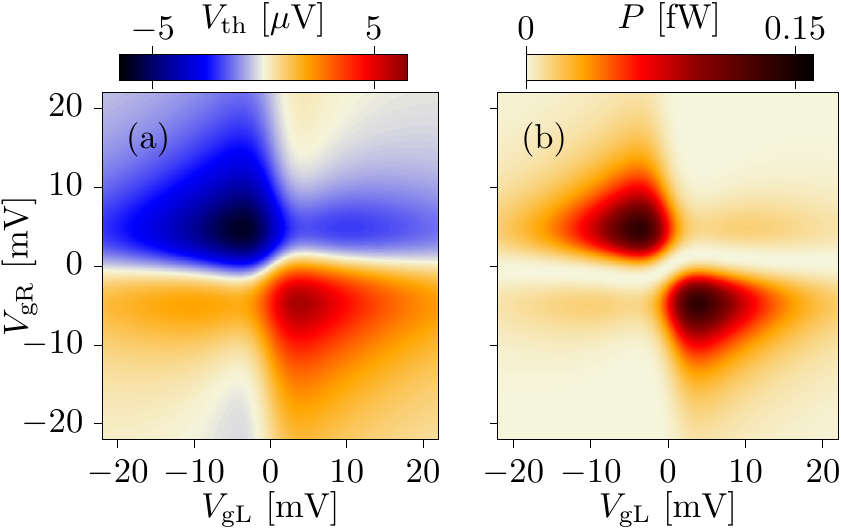}
    \caption{ 
Theoretical calculation of (a) the thermovoltage and (b) the generated power with parameters of $A_{\rm L}=A_{\rm R}=1, \Gamma_{\rm L}=\unit[0.2]{meV}, \Gamma_{\rm R}=\unit[0.1]{meV}$, $\alpha_{\rm L}=0.026$ and  
$\alpha_{\rm R}=0.014$, the base temperature of $T_{\rm 0}=\unit[85]{mK}$ and the cavity temperature of $T_{\rm C}=\unit[170]{mK}$. The influence of the external load resistance is taken into account.}
    \label{theory}
\end{figure}

The experimental data in Fig.~\ref{Vth01} and Fig.~\ref{VandP} is reproduced by the model of Ref.~\cite{jordan_powerful_2013} which we generalize to incorporate the effect of the external circuit. The thermoelectric transport through each dot can be described by the Landauer-Büttiker formalism, with the expression:
\begin{equation}
     {\mathcal I}_{l,n} = \frac{2}{h}\int dE E^n{\cal T}_l(E)[f_l(E)-f_{\rm C}(E)],
     \label{mol1}
\end{equation} 
giving the charge {$I_l=e \mathcal I_{l,0}$} and energy {$J_l=\mathcal I_{l,1}$} currents at lead $l$=L,R.
The quantum dot resonances are defined by a transmission coefficient
\begin{equation}
     {\cal T}_l(E)=A_l\frac{\Gamma^2_l/4}{(E-\varepsilon_l)^2+\Gamma^2_l/4},
     \label{mol2}
\end{equation}
where the parameter $A_l$ depends on the asymmetry of the quantum dot barriers \cite{buttiker_coherent_1988}. The quantum dot resonant energies are tuned with gate voltages, $\varepsilon_l=\varepsilon_{l,0}-e\alpha_lV_{{\rm g}l}$. In our experiment, the width $\Gamma_l$ is thermally broadened beyond the natural line width of the level. As no charge is injected from the heating channel into the conductor, the conservation laws for charge and energy read:
\begin{equation}
I_{\rm L}+I_{\rm R}=0, \qquad J_{\rm L}+J_{\rm R}+{\dot{Q}}=0.
\label{mol4}
\end{equation}
where ${\dot{Q}}$ is the heat current injected into the central cavity. 
For a closed circuit where the energy harvester powers an impedance $R_{\rm Load}$, the voltages are set via Ohm's law, producing the thermovoltage, $V_{\rm th} = I_{\rm L} R_{\rm Load}$ and power of Fig.~\ref{theory}. Accounting for the external resistance in the circuit gives rise to additional features not present in an open-circuit model~\cite{jordan_powerful_2013}, such as the vertical and horizontal lines in Fig.~\ref{theory}. Our simple model based on resonant tunneling captures all the features of the experiment, seen by the comparison of experimental data (points of black circles, red stars and blue triangles) and theoretical modeling results (solid lines in black, red and blue) in Fig.~\ref{VandP}.  
The theoretical efficiency, shown as solid lines in Fig.~\ref{VandP}(c), is computed with the general expression of the heat current evaluated at the obtained thermovoltage, 
\begin{equation}
\label{eq:theff}
\eta=V_{\rm th}^2/(\dot{Q}R_{\rm Load}).
\end{equation}
Fig.~\ref{VandP}(c) suggests the top theoretical efficiency of the device is $\sim0.5 \eta_C$, for the considered parameters. Given its overall good agreement with experimental results, this theoretical model provides a more realistic estimate of the efficiency with its direct access to the heat current, $\dot{Q}$. The experimental estimates in Fig.~\ref{VandP}(c), extracted by Eq.~(\ref{efficiency}), are only the lower bound of the efficiency, where the thermal conductance is overestimated because quantum dots have a smaller Lorenz ratio than $L$ due to the violation of the Wiedemann-Franz law as discussed earlier~\cite{kubala_violation_2008, tsaousidou_thermoelectric_2010,erdman_thermoelectric_2017,dutta_thermal_2017}.

In conclusion, we have experimentally realized an energy harvester based on resonant-tunneling quantum dots~\cite{jordan_powerful_2013} which can generate a power of $\unit[0.13]{fW}$ in an estimated efficiency with a lower bound around $\unit[0.1]{\eta_{\rm C}}$. Our theoretical model (not affected by limitations of the  Wiedemann-Franz law) suggests the actual efficiency to be about $\unit[0.5]{\eta_{\rm C}}$. Experimental observations of thermal power, voltage and efficiency at different values of $I_{\rm Heat}$ and $R_{\rm Load}$ have also been reproduced by this model. There are small quantitative differences between experimental results and theoretical modeling in terms of parameters, such as electrical temperatures and energy level difference. This may be explained by asymmetric barriers, accidental degeneracies or the broadened lifetime width of the quantum dots, as well as charging effects in the non-linear regime. Also, the oscillation brought with the AC heating and AC measurements can increase thermal broadening in the cavity, and therefore cause inaccuracy in the measurement results. Overall, this proof-of-principle experiment demonstrates the basic soundness of the theory of mesoscopic energy harvesting with energy filtering techniques at the quantum level, realizing a heat engine. 

We propose several possible improvements for future work. First, we can improve the power and efficiency by optimizing the resonance width $\Gamma_{l}$ as well as the symmetry of the quantum dots. Second, DC heating and measurement techniques can be used to avoid unnecessary oscillations of voltages and temperatures in the device. Finally, the performance of the energy harvester may be enhanced by scaling it up in size with resonant-tunneling quantum wells, which may increase the maximum power up to fractions of W/cm$^{2}$ at \unit[300]{K}~\cite{sothmann_powerful_2013}, or by using smaller dots or molecules, whose large level spacing allows the system to operate at higher temperatures~\cite{leobandung_observation_1995,park_coulomb_2002,jordan_powerful_2013}. 

We are grateful to Dr. J. Waldie for technical assistance. This work was funded by EPSRC(UK). GJ acknowledges financial support from China Scholarship Council and GBCET. RS acknowledges financial support from the Spanish MINECO via grant FIS2015-74472-JIN (AEI/FEDER/UE), the Ram\'on y Cajal program RYC-2016-20778 and through the ``María de Maeztu” Programme
for  Units  of  Excellence  in  R\&D  (MDM-2014-0377). Work by ANJ was supported by the U.S. Department of Energy (DOE), Office of Science, Basic Energy Sciences (BES) under Award \#DE-SC0017890. 
BS acknowledges financial support from the Ministry of Innovation NRW via the ``Programm zur Förderung der Rückkehr des hochqualifizierten Forschungsnachwuchses aus dem Ausland''. This research was supported in part by the National Science Foundation under Grant No. NSF PHY-1748958. 

\bibliographystyle{apsrev4-1}
\bibliography{Meine_Bibliothek}

\clearpage

\pagebreak

\section{Supplementary Material:  Experimental realisation of a quantum dot energy harvester } \label{supplementary}

\subsection{Estimating temperature differences}

\begin{figure}[b!]
    \centering    \includegraphics[width=0.95\linewidth]{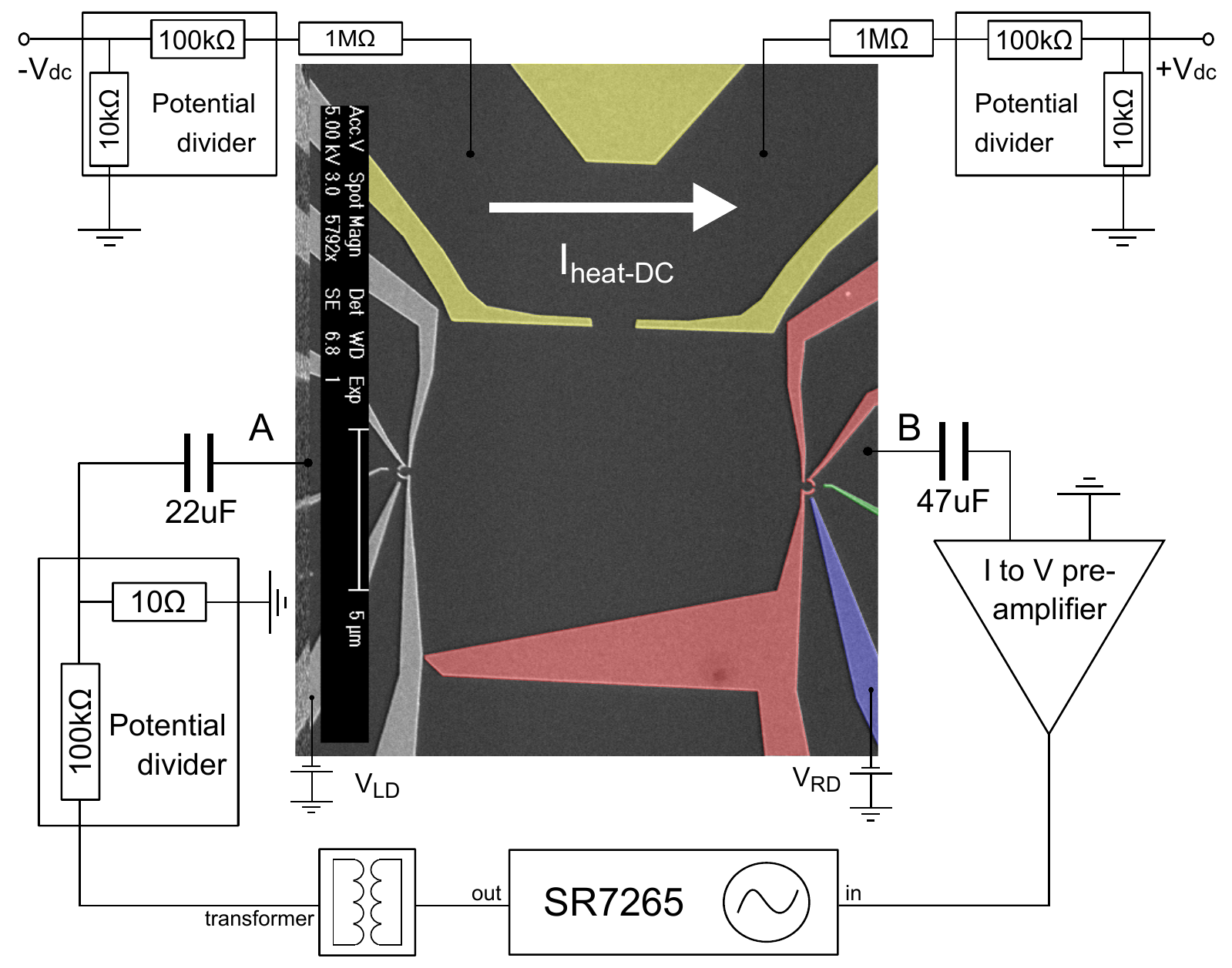}
    \caption{False colour SEM image of the device with the electrical circuit used for the temperature measurement.}
    \label{T-MeasurementSet-up}
\end{figure}

\begin{figure}[tb!]
    \centering    \includegraphics[width=0.85\linewidth]{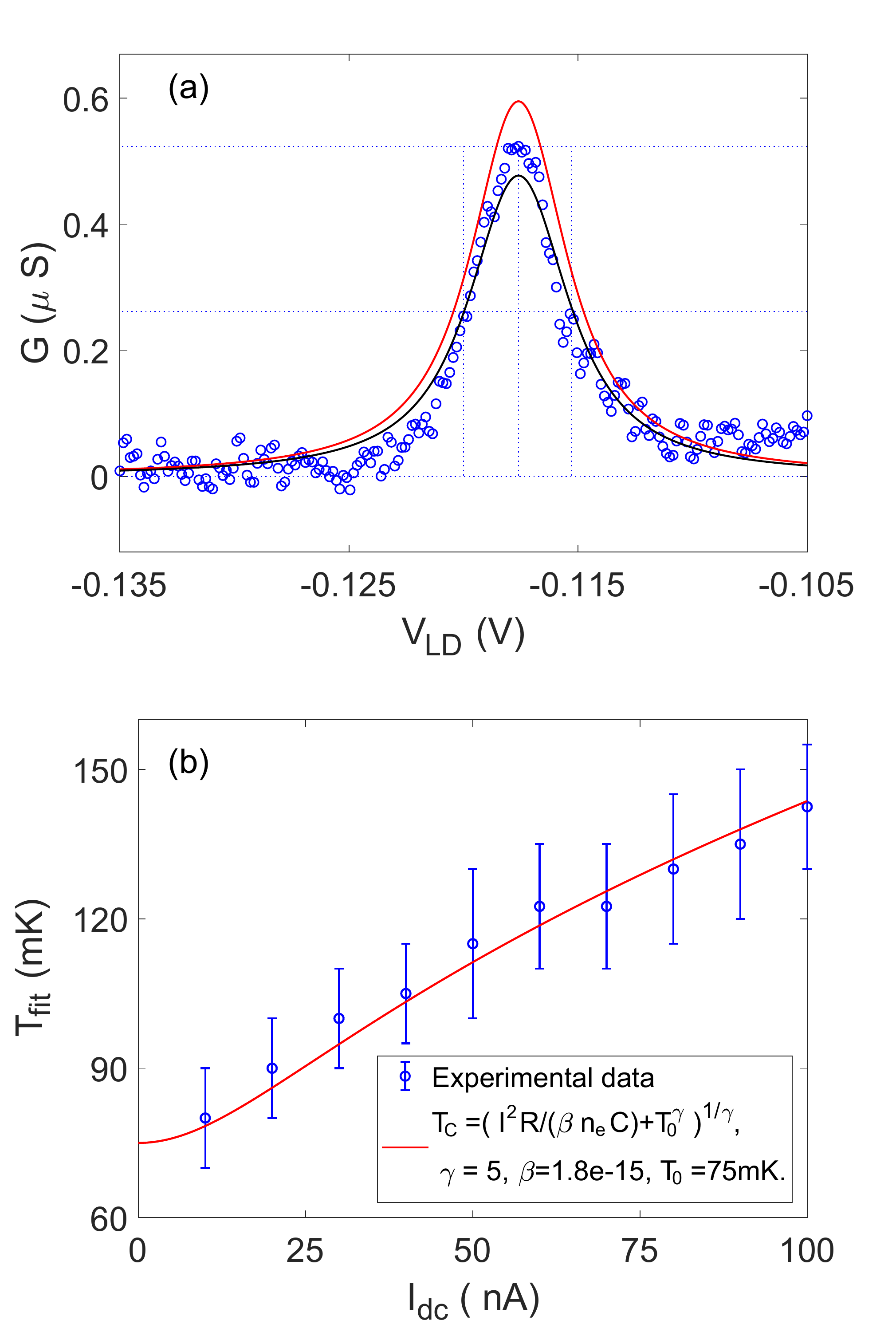}
    \caption{(a) Differential conductance $G$ plotted versus plunger voltage $V_\text{LD}$. The plot is a representative Coulomb resonance peak, measured at a lattice temperature of \unit[75]{mK} and with a DC heating current of \unit[30]{nA} applied to the heating channel. The circles are the experimental data points and the red and the black lines are the minimum \unit[90]{mK} and maximum \unit[110]{mK} theoretical fits by using Eq.~(\ref{tempequ}). (b) The blue dots are the estimated electrical temperatures $T_C$ using Eq.~(\ref{tempequ}), with the fitting method shown in (a), and the red line is the relative general relationship between the energy loss rate and the temperature. }
    \label{Tfit}
\end{figure}

In this section, we discuss how the temperatures in Fig.~\ref{model} (b) are estimated. Figure \ref{T-MeasurementSet-up} shows the circuit used for temperature measurements. Two potential dividers of 1/10th are set on either side of the heating channel, followed by a \unit[1]{M$\Omega$} resistor, to provide constant currents in the channel. DC voltages of the same magnitude but opposite sign are applied spontaneously to either side of the heating channel via {\em NicDAQ-9178}, to avoid a bias voltage being applied across the quantum dots. Two capacitors of \unit[22]{$\mu$F} and \unit[47]{$\mu$F} are set on either side of the cavity, to block heating currents from leaking through the dots. Therefore, heating currents only go to the ground of the potential divider on the other side the heating channel. The differential conductance, $G=dI/dV$, are measured through the left quantum dot while different DC heating currents are applied in the channel, and the cavity is defined. To extract the temperature of the cavity at each DC heating current, $G$ is fitted to a thermally broadened Lorentzian, parameterised by a full width at half-maximum $\Gamma$ \cite{foxman_effects_1993},

\begin{equation}\label{tempequ}
\begin{aligned}
 G = \frac{e^2}{h} \frac{1}{4k_{B}T_C}&B\int_{-\infty}^{+\infty} \cosh^{-2}\left(\frac{E}{2k_{B}T_C} \right) \times \\
       &\frac{(\Gamma/2)\pi}{(\Gamma/2)^2 +[(e\alpha V_g-E_{res})-E]^2} dE. 
\end{aligned}
\end{equation}
\noindent
Where $e$ is the electron charge, $k_{B}$ is the Boltzmann Constant, $T_\text{C}$ is the electron temperature of the cavity, $B$ is the temperature-independent energy-integrated strength of the resonance, $E$ and $E_{res}$ are the energy of the dot level and the energy of resonance respectively. Here $E=e \alpha V_{LD}$, with $\alpha = 0.025$ the lever arm of the plunger gate, ascertained via measurement of the Coulomb Diamonds for each dot, which is discussed in the next section. For DC currents from $\unit[0]{}$ to $\unit[100]{nA}$,  Eq.~\eqref{tempequ} gives a temperature range from $\unit[75]{mK}$ to $\unit[150]{mK}$, as shown in Fig.~\ref{Tfit}(b). 

Our data shows the power dissipated per electron \cite{appleyard_thermometer_1998} is best fit with $P=I^2R/(n_eC)=\beta (T_C^5-T_0^5)$ with $R=\unit[500]{\Omega}$, $C=\unit[350]{\mu m^2}$ and $\beta = \unit[1.8 \times 10^{-15}]{WK^{-5}}$, shown as the red line in Fig.~\ref{Tfit}(b). Here, $I$ is the heating current, $R$ is the resistance of the heating channel and $C$ is the area of the heating channel. This $T^5$ behavior was attributed to acoustic phonon scattering in the Gr\"{u}neisen-Bloch regime in the heating channel, with coupling via a screened piezoelectric potential \cite{proskuryakov_energy-loss_2007}. The $\beta = \unit[1.8\times 10^{-15}]{WK^{-5}}$ is larger than the theoretical prediction of $\unit[9.6\times 10^{-18}]{ WK^{-5}}$ ($\unit[60]{ eV/sK^5}$) \cite{appleyard_thermometer_1998}. This can be because some heat is leaking through ohmic contacts at each end of the heating channel. For $T\sim \unit[500]{mK}$, the heated electrons relax to the lattice temperature over a distance $l\textsubscript{e-ph}\sim \unit[200]{\mu m}$, and as $T$ is lowered further, $l\textsubscript{e-ph}$ can significantly exceed the size of the cavity of the device \cite{dzurak_thermoelectric_1997}. In this regime, the energy redistribution is achieved via electron diffusion to the cold reservoirs \citep{mittal_electron-phonon_1996}. Therefore, in this device, hot electrons diffuse out from the heating channel to the central cavity and replace cold electrons, serving to redistribute energy and leading to a well-defined electron temperature profile.   

We assume the central cavity and the reservoir of the heating channel share the same temperature $T_C$, and the left and the right reservoirs have the same temperature with the base temperature, $T_L=T_R=T_0$.
The base temperature estimated by this analysis is $\unit[75]{mK}$. The quantum dot was used as the thermometer in this experiment, because it has greater accuracy than the thermometer in the mixing chamber of this dilution refrigerator, which gave a base temperature of $\unit[50]{mK}$. 

\subsection{Coulomb blockade analysis}

\begin{figure}[b]
    \centering
    \includegraphics[width=0.85\linewidth]{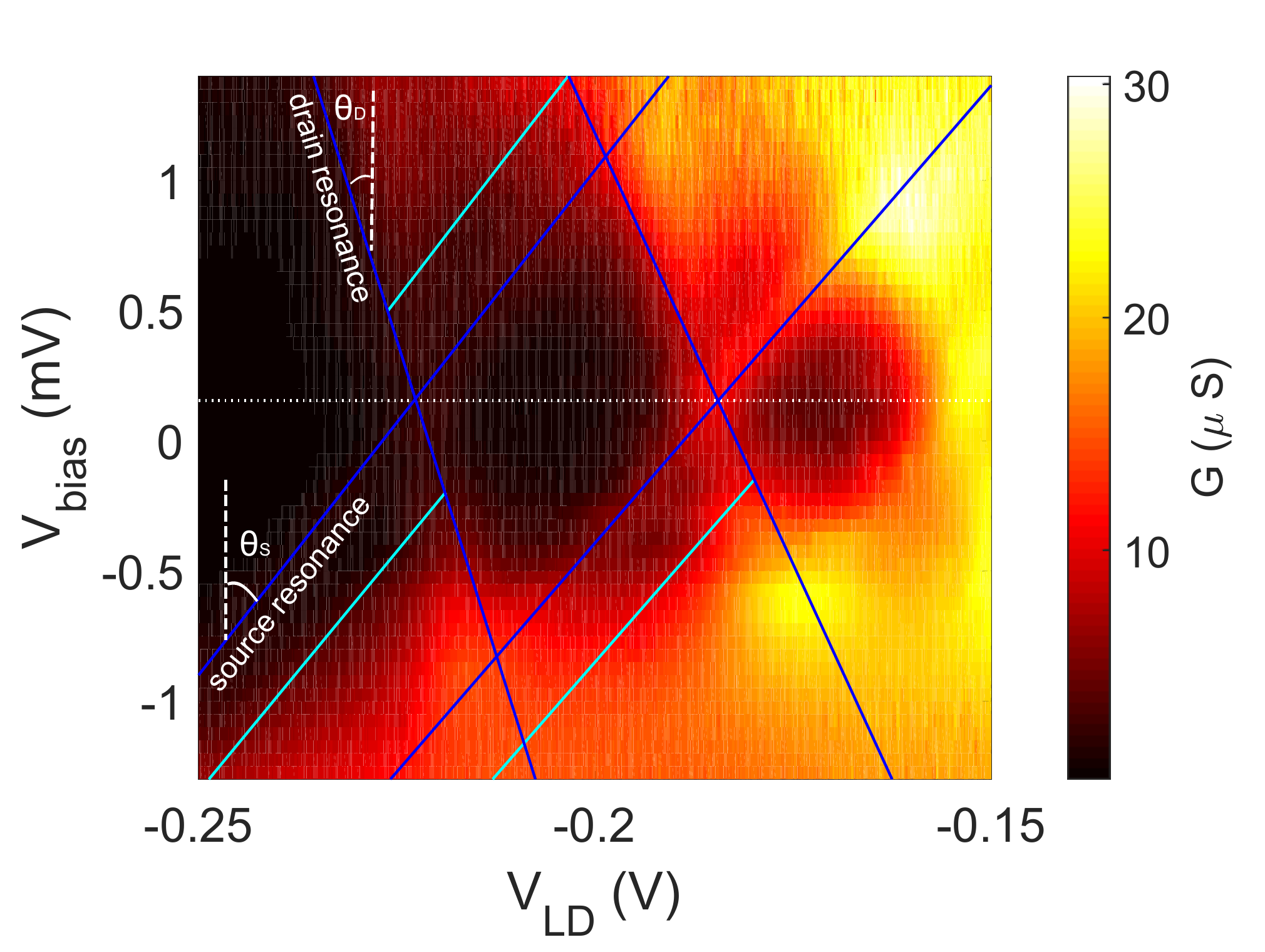}
    \caption{Coulour scale plot of measured transport through the left quantum dot in Fig.~\ref{model}, plotted as a function of the voltage on the plunger gate V\textsubscript{LD}, and of the source drain bias V\textsubscript{bias}. The 'Coulomb Diamond' in which transport through the dot is blockaded is depicted by the blue solid lines. Increasing the source drain bias increases the range of plunger voltage over which the dot is unblocked. At a sufficiently large source drain bias, the first excited level enters the transport window, and conduction through the dot increases, depicted as the light blue lines in the figure. The horizontal dashed line marks the actual zero bias point of Coulomb Diamonds, and the V\textsubscript{bias}=\unit[0.155]{mV} is the offset from the measurement. }
    \label{Diamond}
\end{figure} 

When an isolated island of charges have a sufficiently small capacitance, the energy required to change its charge by even one electron may be large. Until this energy is supplied, no charge may move onto or off the island. This is called \textit{Coulomb blockade}. During our experiment, the two quantum dots have firstly been tuned to have energy levels lying between the potentials of source and drain reservoirs. In this situation, the Coulomb blockade has been lifted and electrons with energies that match the energy level lying within the bias window can tunnel through the dots. When the potential of the dots is changed by a gate electrode, the plunger gate in Fig.~\ref{model}, the current through a quantum dot shows periodic oscillations in plunger-gate voltage, which is known as the \textit{Coulomb Peak}. Increasing the source-drain bias across a quantum dot will increase the window of energies over which states in the source are full and states in the drain are empty. With a large enough bias, the widened conductance peaks from adjacent charge states of dots overlap leaving the diamond shape regions, which are commonly referred to as \textit{Coulomb Diamonds} \cite{prance_electronic_2009}. In reality, the excited states start contributing to transport before the transport window is wide enough to include the next charge state of the dot, shown as light blue lines in Fig.~\ref{Diamond}.
The boundaries of the Coulomb diamonds are labeled as `source resonance' and `drain source' in Fig.~\ref{Diamond}. This is because they correspond to the dot level being aligned with the source and drain potentials respectively. The gradient of these resonances can be used to calibrate the conversion factor between $\Delta V\textsubscript{g}$ and the electrochemical potential of the dot. The gradient are defined as \cite{prance_electronic_2009}:

\begin{equation}
     m\textsubscript{S}= \tan(\theta \textsubscript{S}) = \left(\frac{\Delta V\textsubscript{g}}{\Delta V\textsubscript{SD}}\right)^{(S)},
     \label{capa2} 
\end{equation}
\begin{equation}
     m\textsubscript{D}= \tan(\theta \textsubscript{D}) = \left(\frac{\Delta V\textsubscript{g}}{\Delta V\textsubscript{SD}}\right)^{(D)}.
     \label{capa3}
\end{equation}
The superscript in the right hand expression denotes whether the gradient is of the source or drain resonance line. The gate electrode \textit{lever arm} is given by the value as: 
\begin{equation}
     \alpha \textsubscript{G}=\frac{1}{m\textsubscript{S}- m\textsubscript{D}} . \label{capa4}
\end{equation}
The \textit{lever arm} for the gating effect on the dot energy from the biased reservoir (the drain) can also be found:
\begin{equation}
 \alpha \textsubscript{D}=\frac{1}{1-m\textsubscript{S}/ m\textsubscript{D}}   \label{capa5}.
\end{equation}\vspace{2mm}
With the bias applied to only the drain reservoir, $\alpha \textsubscript{S}$, the \textit{lever arm} of the source cannot be found. 

The lever arm of a gate (or lead) quantifies the effect of the gate on the potential of the dot. It is a useful parameter to convert the experimentally measured voltage to energy. During this experiment, Coulomb diamonds from the left dot gives 0.025 lever arm, which is crucial to the analysis of electrical temperatures.

\end{document}